\documentclass[amsmath,amssymb,twocolumn,prd,floatfix,showpacs, nofootinbib]{revtex4-2}
\usepackage{graphicx}
\usepackage{tabularx}  
\usepackage{bm, color} 
\usepackage{overpic,subfigure} 
\usepackage{multirow}
\usepackage{array}
\usepackage{dcolumn} 
\usepackage[symbol]{footmisc}
\usepackage{booktabs}
\usepackage{epstopdf}
\usepackage[normalem]{ulem}
\usepackage{hyperref}
\RequirePackage{xspace}
\newcommand{\gev}{\ensuremath{\mathrm{\,Ge\kern -0.1em V}}\xspace}
\newcommand{\mev}{\ensuremath{\mathrm{\,Me\kern -0.1em V}}\xspace}
\newcommand{\mevcc}{\ensuremath{{\mathrm{\,Me\kern -0.1em V\!/}c^2}}\xspace}

\def\fz#1       {\ensuremath{f_0({#1})}\xspace}

\usepackage{lineno}
\hypersetup{colorlinks = true,
            linkcolor = blue,
            urlcolor = blue,
            citecolor = blue,
            breaklinks=true,
            pdfstartview=Fit}
\begin{document}


\newcommand{\BESIIIorcid}[1]{\href{https://orcid.org/#1}{\hspace*{0.1em}\raisebox{-0.45ex}{\includegraphics[width=1em]{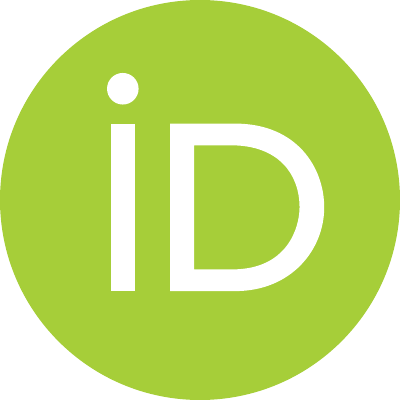}}}}

\title{\boldmath Improved amplitude analysis of $\eta^\prime\to\pi^+\pi^-\pi^0$ and $\eta^\prime\to\pi^0\pi^0\pi^0$ }

\author{
\begin{center}
M.~Ablikim$^{1}$\BESIIIorcid{0000-0002-3935-619X},
M.~N.~Achasov$^{4,c}$\BESIIIorcid{0000-0002-9400-8622},
P.~Adlarson$^{84}$\BESIIIorcid{0000-0001-6280-3851},
X.~C.~Ai$^{90}$\BESIIIorcid{0000-0003-3856-2415},
C.~S.~Akondi$^{32A,32B}$\BESIIIorcid{0000-0001-6303-5217},
R.~Aliberti$^{40}$\BESIIIorcid{0000-0003-3500-4012},
A.~Amoroso$^{83A,83C}$\BESIIIorcid{0000-0002-3095-8610},
Q.~An$^{79,66,\dagger}$,
M.~S.~Anderson$^{40}$\BESIIIorcid{0009-0008-1550-2632},
Y.~Bai$^{64}$\BESIIIorcid{0000-0001-6593-5665}
% ...
\\
\vspace{0.2cm}
(BESIII Collaboration)\\
\vspace{0.2cm} {\it
$^{1}$ Institute of High Energy Physics, Beijing 100049, People's Republic of China\\
% ...
$^{4}$ Budker Institute of Nuclear Physics SB RAS (BINP), Novosibirsk 630090, Russia\\
% ...
$^{32}$ INFN Sezione di Ferrara, (A)INFN Sezione di Ferrara, I-44122, Ferrara, Italy; (B)University of Ferrara, I-44122, Ferrara, Italy\\
% ...
$^{40}$ Johannes Gutenberg University of Mainz, Johann-Joachim-Becher-Weg 45, D-55099 Mainz, Germany\\
% ...
$^{64}$ Southeast University, Nanjing 211100, People's Republic of China\\
% ...
$^{66}$ State Key Laboratory of Particle Detection and Electronics, Beijing 100049, Hefei 230026, People's Republic of China\\
% ...
$^{79}$ University of Science and Technology of China, Hefei 230026, People's Republic of China\\
% ...
$^{83}$ University of Turin and INFN, (A)University of Turin, I-10125, Turin, Italy; (B)University of Eastern Piedmont, I-15121, Alessandria, Italy; (C)INFN, I-10125, Turin, Italy\\
$^{84}$ Uppsala University, Box 516, SE-75120 Uppsala, Sweden\\
% ...
$^{90}$ Zhengzhou University, Zhengzhou 450001, People's Republic of China\\

\vspace{0.2cm}
$^{\dagger}$ Deceased\\
% ...
$^{c}$ Also at the Novosibirsk State University, Novosibirsk, 630090, Russia\\
% ...

}

\end{center}
}

\begin{abstract}

Using a sample of $(10087\pm44)\times 10^6$ $J/\psi$ events collected with the BESIII detector at BEPCII, we perform an amplitude analysis of the decays $\eta^\prime\to\pi^+\pi^-\pi^0$ and $\eta^\prime\to\pi^0\pi^0\pi^0$, where we observe significant $\pi^\pm\pi^0$ $P$-wave and $\pi$-$\pi$ $S$-wave interactions. Two different parameterizations, a $\pi$-$\pi$ scattering phase shift and the Gounaris–Sakurai Breit–Wigner formalism, are used to describe the $P$-wave propagator.
Due to the large interference, the branching fractions for both the $P$- and the $S$-waves are found to be strongly model dependent.
\end{abstract}

\maketitle

\section{Introduction}

The isospin-violating decays $\eta/\eta^\prime\to\pi\pi\pi$ are induced dominantly by the strong interaction via the explicit breaking of chiral symmetry by the $d-u$ quark mass difference, which has garnered widespread attention both theoretically~\cite{Gross:1979ur, Borasoy:2005du, Borasoy:2006uv} and experimentally~\cite{ref::CLEO, ref::BESetap, ref::kang}. 
Reference~\cite{Gross:1979ur} pointed out that the light quark mass difference can be extracted with the ratios between decay widths of $\eta^\prime\to\pi\pi\pi$ and $\eta^\prime\to\eta\pi\pi$, under the two assumptions: (a) the decay $\eta^\prime\to\pi\pi\pi$ proceeds entirely via the decay $\eta^\prime\to\eta\pi\pi$ followed by $\pi^0$-$\eta$ mixing and (b) the decay amplitudes are constant over phase space.
However, such assumptions have proven not to be fulfilled by a measurement of the \text{BESIII} Collaboration performed in 2017~\cite{ref::kang}. 
Indeed, a large contribution from $P$-wave intermediate states has been observed in the decay $\eta^\prime \to \pi^+\pi^-\pi^0$, while this contribution is forbidden for the $3\pi^0$ final state due to Bose symmetry. 
Moreover, a resonant $\pi$-$\pi$ $S$-wave contribution, interpreted as the broad $\sigma$ meson, has been found to play an essential role in both the charged and neutral decay modes. 
Finally, evidence for $\pi$-$\pi$ scattering has been observed in $\eta^\prime \to \eta\pi^0\pi^0$~\cite{BESIII:2022tas}, which corresponds to the cusp effect predicted by the non-relativistic effective field theory~\cite{Kubis:2009sb}. 
Therefore, considerable theoretical improvements are still required before determining the light quark mass difference from the $\eta^\prime \to \pi\pi\pi$ decays.

Recently, the branching fraction of the decay $\eta^\prime\to\rho^\pm\pi^\mp$ was calculated to be $(5.3\pm0.9)\times10^{-4}$ in the chiral quark Nambu-Jona-Lasinio (NJL) model~\cite{Volkov:2022nuf}. This is consistent with the \text{BESIII} measurement,
$(7.44 \pm 0.60_{\rm{stat.}}\pm1.26_{\rm{syst.}}\pm 1.84_{\rm{model}}) \times 10^{-4}$, where the first uncertainty is statistical, the second systematic, 
and the third is from the $\rho$-propagator model. 
The systematic and propagator-model uncertainties are relatively large and dominate the total uncertainty.
With the additional $J/\psi$ data collected in 2018 and 2019, the total number of $J/\psi$ events accumulated with the BESIII detector is increased to $(10087 \pm 44)\times10^6$~\cite{ref::Jpsi}, which is about eight times the sample used in the previous BESIII analysis. In addition to improving the statistical precision, the larger dataset enables better control of background contributions and efficiency-related systematic uncertainties,  
thereby enabling a more precise amplitude analysis of the decays of $\eta' \to \pi^+\pi^-\pi^0$ and $\eta' \to \pi^0\pi^0\pi^0$.

\section{BESIII DETECTOR AND MONTE CARLO SIMULATION}

The \text{BESIII} detector~\cite{Ablikim:2009aa} records symmetric $e^+e^-$ collisions 
provided by the BEPCII storage ring~\cite{Yu:IPAC2016-TUYA01}
in the center-of-mass energy range from 1.84 to 4.95~GeV, with a peak luminosity of $1.1 \times 10^{33}\;\text{cm}^{-2}\text{s}^{-1}$ 
achieved at $\sqrt{s} = 3.773\;\text{GeV}$. \text{BESIII} has collected large data samples in this energy region~\cite{Ablikim:2019hff,EcmsMea,EventFilter}. The cylindrical core of the \text{BESIII} detector covers $93\%$ of the full solid angle and consists of a helium-based
 multilayer drift chamber~(MDC), a time-of-flight
system~(TOF), and a CsI(Tl) electromagnetic calorimeter~(EMC),
which are all enclosed in a superconducting solenoidal magnet
providing a 1.0~T magnetic field. The magnetic field was 0.9~T in 2012, which affects $11\%$ of the total $J/\psi$ data. The solenoid is supported by an
octagonal flux-return yoke with resistive plate counter muon
identification modules interleaved with steel. 
The charged-particle momentum resolution at $1~{\rm GeV}/c$ is
$0.5\%$, and the 
${\rm d}E/{\rm d}x$
resolution is $6\%$ for electrons
from Bhabha scattering. The EMC measures photon energies with a
resolution of $2.5\%$ ($5\%$) at $1$~GeV in the barrel (end cap)
region. The time resolution in the plastic scintillator TOF barrel region is 68~ps, while
that in the end cap region was 110~ps. The end cap TOF
system was upgraded in 2015 using multigap resistive plate chamber
technology, providing a time resolution of
60~ps,
which benefits $87\%$ of the data used in this analysis~\cite{Li:2017etof,Guo:2017etof,Cao:2020etof}.

Monte Carlo (MC) simulated data samples produced with a {\sc
geant4}-based~\cite{geant4} software package, which
includes the geometric description of the \text{BESIII} detector and the
detector response, are used to determine detection efficiencies
and to estimate backgrounds. The simulation models the beam
energy spread and initial state radiation in the $e^+e^-$
annihilations with the generator {\sc
kkmc}~\cite{kkmc1,kkmc2}. All particle decays are modelled with {\sc
evtgen}~\cite{evtgen1,evtgen2} using branching fractions 
either taken from the
Particle Data Group (PDG)~\cite{pdg}, when available,
or otherwise estimated with {\sc lundcharm}~\cite{lundcharm1,lundcharm2}.
Final state radiation
from charged final state particles is incorporated using the {\sc
photos} package~\cite{photos2}.

\section{event selection and background analysis of the decay $\eta^\prime\to\pi^+\pi^-\pi^0$ }

To reconstruct $J/\psi\to\gamma\eta'$ with $\eta^\prime\to\pi^+\pi^-\pi^0$ and $\pi^0\to\gamma\gamma$, candidate events are required to have two oppositely charged tracks and at least three photon candidates. Charged tracks detected in the MDC are required to be within a polar angle range of $|\rm{cos}\theta|<0.93$. The distance of closest approach to the interaction point (IP) must be within 10 cm along the beam axis and 1 cm in the perpendicular plane. Photon candidates are identified using isolated showers in the EMC. The deposited energy of each shower is required to be larger than 25~MeV in the barrel region $|\rm{cos}\theta|<0.8$ and 50~MeV in the endcap regions ($ 0.86 < |\rm{cos}\theta|<0.92$). To exclude showers that originate from charged tracks, the angle subtended by the EMC shower and the position of the closest charged track at the EMC must be greater than 10 degrees as measured from the IP. To suppress electronic noise and showers unrelated to the event, the difference between the EMC time and the event start time determined from charged tracks is required to be within [0, 700]\,ns.

Since the radiative photon ($\gamma_R$) from the $J/\psi$ decay is always more energetic than the photons from the $\pi^0$ decay, the photon candidate with the maximum energy is  
taken to be the radiative photon $\gamma_R$ 
from the $J/\psi$ decay. 
For each $\pi^+\pi^-\gamma\gamma\gamma$ combination, a six-constraint (6C) kinematic fit is performed and the $\chi^2_{\rm{6C}}(3\gamma)$ is required to be less than 25, imposing energy-momentum conservation and constraining the invariant masses of $\gamma\gamma$ and $\pi^+\pi^-\pi^0$ combinations to the nominal $\pi^0$ and $\eta^\prime$ mesons, respectively. 
For events with more than three photons, the combination with the smallest 
$\chi^2_{\rm{6C}}(3\gamma)$ is retained. 
To suppress potential background events with two or four photons in the final state, we further require that the probability of the 6C kinematic fit for the signal hypothesis is greater than the probability of the 4C kinematic fit imposing energy-momentum conservation for the $J/\psi\to\pi^+\pi^-2\gamma$ and $J/\psi\to\pi^+\pi^-4\gamma$ background hypotheses.

\begin{figure}[htbp]
  \centering
  \includegraphics[width=0.48\textwidth]{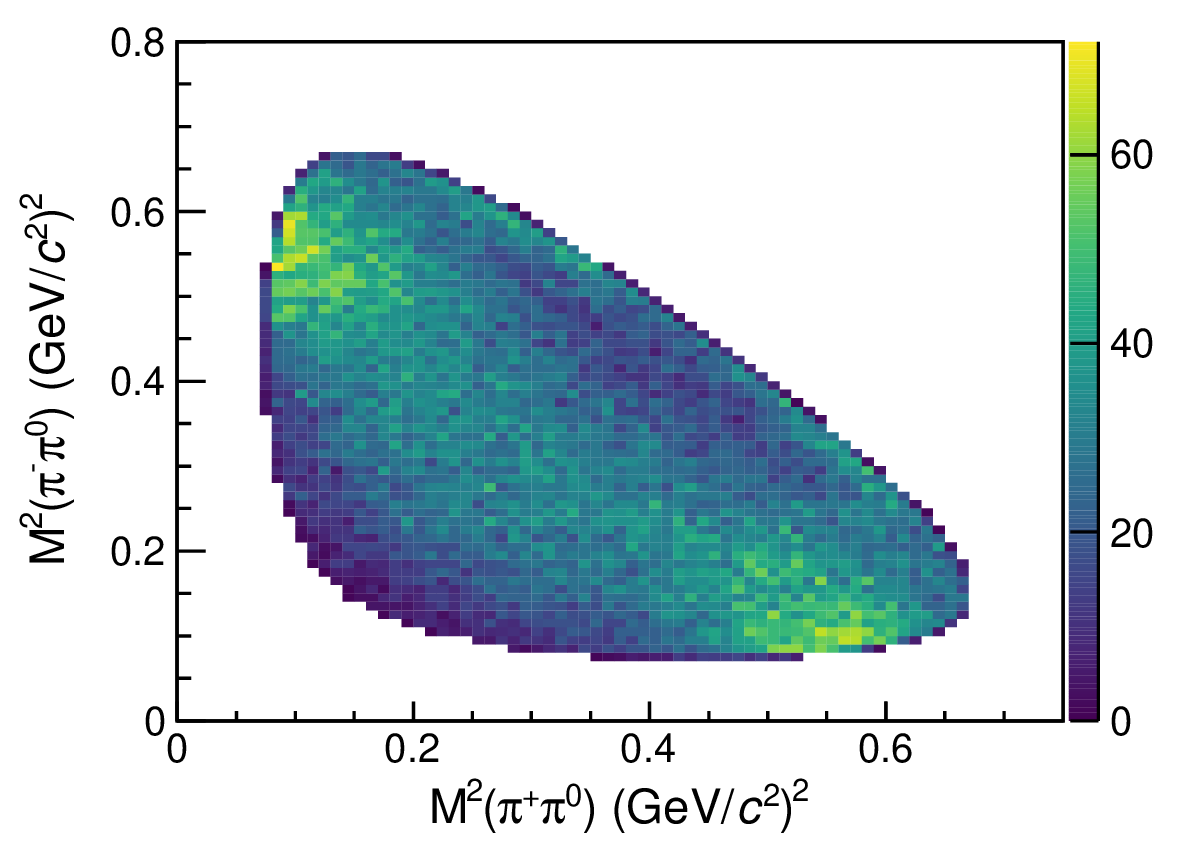}
  \vspace{-0.25cm}
  \caption{
     Dalitz plot for the $\eta^\prime\to \pi^+\pi^-\pi^0$ candidate events selected from data. 
  }
  \label{fig:dalitz_mass}
\end{figure}

To reduce backgrounds from incorrect $\pi^0$ combinations, the invariant mass of the $\gamma_R$ from $J/\psi$ boson and one $\gamma$ from the $\pi^0$ meson are required to be outside of the $\pi^0$ signal region ($|M(\gamma_{R}\gamma_{\pi^{0}})-M_{\pi^0}|>0.016$~ GeV$/c^{2}$).
Furthermore, events with $|M(\gamma\pi^{0})-M_{\omega}|>0.05$~ GeV$/c^{2}$ are rejected to suppress background from the $J/\psi\to\omega\pi^+\pi^-$ decay.
A sample of 65,676 candidate events is selected and the corresponding Dalitz plot is shown in Fig.~\ref{fig:dalitz_mass}. Two clear clusters corresponding to the decays of $\eta^\prime\to\rho^{\pm}\pi^{\mp}$ are observed, as expected.

The possible background sources are investigated with a MC sample of about 10 billion $J/\psi$ inclusive events. Using the same selection criteria, the surviving background events mainly originate from the decays $\eta^\prime\to\gamma\pi^+\pi^-$ and $\eta^\prime\to\gamma\rho^0$ with $\rho^0\to\gamma\pi^+\pi^-$, which show up as peaks around the $\eta^\prime$ mass region, and the non-peaking process $J/\psi\to\pi^+\pi^-\pi^0\pi^0$ with multiple photons in the final state. 
For the $\eta^\prime\to\gamma\pi^+\pi^-$ decay, a dedicated MC simulation based on the \text{BESIII} amplitude analysis~\cite{ref::chapeak} is generated, including the contributions from the $\rho^0$, $\omega$ resonances as well as the box anomaly~\cite{box1,box2}. 
For the $\eta^\prime\to\gamma\rho^0$ with $\rho^0\to\gamma\pi^+\pi^-$ decay, an MC sample is generated based on the amplitude in Ref.~\cite{ref::rhogpp}. 
Using the corresponding branching fractions from the PDG, the expected numbers of events for $\eta^\prime\to\gamma\pi^+\pi^-$ and $\eta^\prime\to\gamma\rho^0\to\gamma\gamma\pi^+\pi^-$ decays are estimated to be $15,110\pm242$ and $4,300\pm69$, respectively.
A phase space distributed $J/\psi \to \pi^+\pi^-\pi^0\pi^0$ sample is produced to describe the non-peaking background.

\begin{figure}[htbp]
  \centering
  \includegraphics[width=0.48\textwidth]{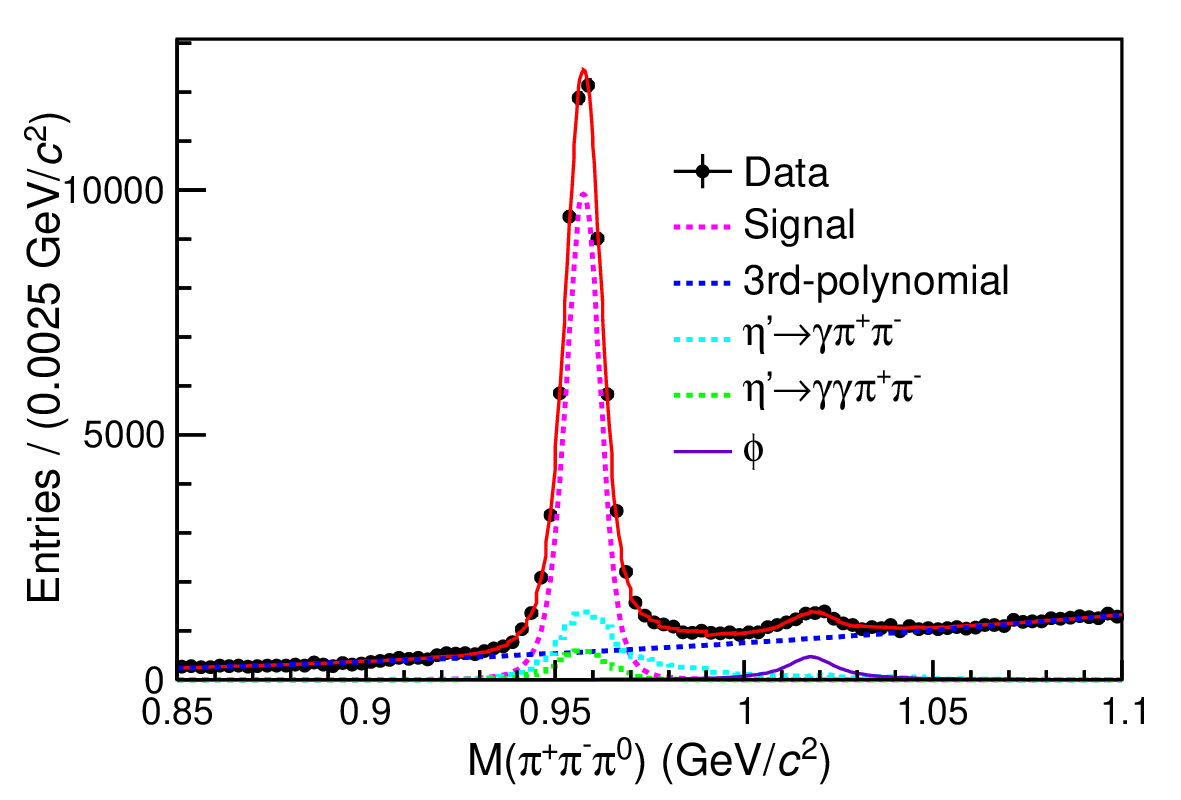}
  \vspace{-0.25cm}
  \caption{
     Invariant mass distribution of $\pi^+\pi^-\pi^0$ candidates without the $\eta^\prime$ mass constraint applied in the kinematic fit. 
  }
  \label{fig:fit_mass}
\end{figure}

To verify that the above MC simulations describe the background well, a data-based validation is performed. An alternative data sample is selected by using a five-constraint (5C) kinematic fit without the $\eta^\prime$ mass constraint. The resulting distribution of the 
 $\pi^+\pi^-\pi^0$ invariant mass, $M(\pi^+\pi^-\pi^0)$, is shown in Fig.~\ref{fig:fit_mass}. 
 An unbinned maximum likelihood fit is performed to the $M(\pi^+\pi^-\pi^0)$ distribution. 
The signal is described by the MC simulated shape convoluted with a Gaussian resolution function, and the peaking backgrounds $\eta'\to\gamma\pi^+\pi^-$ and $\eta'\to\gamma\rho^0\to\gamma\gamma\pi^+\pi^-$ decays are described by the MC simulated shapes with their numbers of events fixed to the expected values, respectively. The non-peaking background is described by a third-order polynomial function, while the small peak around 1.02~ GeV/$c^2$ from the $J/\psi \rightarrow \gamma\gamma\phi$ decay is modeled with a Breit-Wigner function. Based on the number of background events in the 5C-fitted sample under the $\eta^\prime$ signal region ($|M(\pi^+\pi^-\pi^0)-M_{\eta^\prime}|<0.02$ GeV/$c^2$) and taking into account the difference of signal efficiency between 5C and 6C kinematic requirements, the number of non-peaking background events in the selected 6C-fitted sample is estimated to be $6,366 \pm 82$.

To check the background distributions in the $\pi\pi$ mass spectra, a series of fits to the $M(\pi^+\pi^-\pi^0)$ invariant mass distribution are performed in different $\pi\pi$ mass intervals. In this way, we can extract the data-based background shapes, which are consistent with the MC simulated shapes including the contributions from $\eta^\prime\to\gamma\pi^+\pi^-$, $\eta^\prime\to\gamma\rho^0\to\gamma\gamma\pi^+\pi^-$, and $J/\psi \to \pi^+\pi^-\pi^0\pi^0$.

\section{event selection and background analysis of the decay $\eta^\prime\to\pi^0\pi^0\pi^0$ }

To reconstruct $J/\psi\to\gamma\eta^\prime$ with $\eta^\prime\to\pi^0\pi^0\pi^0$ and $\pi^0\to\gamma\gamma$, candidate events are required to have at least seven photons and no charged track. The selection criteria for the photon candidates are the same as described before, except for the requirement on the angle between photon candidates and charged tracks. 
To suppress electronic noise and clusters unrelated to the event, the difference between the EMC time and that of the most energetic photon is required to be within $[-500,500]$ ns. The photon with maximum energy in the event is assumed to be $\gamma_R$ originating from the $J/\psi$ decay.
For the remaining photon candidates, all possible $\gamma\gamma$ pairs are combined into $\pi^0$ candidates and subjected separately to a one-constraint kinematic fit in which the invariant mass of the $\gamma\gamma$ pair is constrained to the nominal $\pi^0$ mass.
The $\chi^2_{\pi^0}$ value of this kinematic fit is required to be less than 25. To suppress photon mis-combinations, the polar angle of the $\pi^0$ decay in the $\pi^0$ helicity frame, defined as $|\cos \theta_h|= |E_{\gamma1} -E_{\gamma2}|/p_{\pi^0}$, where $E_{\gamma1}$, $E_{\gamma2}$ and $p_{\pi^0}$ are
the photon energies and the $\pi^0$ momentum in the lab frame,
respectively, is required to satisfy $|\cos \theta_{h}|<0.95$.
Events with at least three accepted $\pi^0$ candidates are
retained for further analysis.

An eight-constraint (8C) kinematic fit is performed, enforcing the energy-momentum conservation and constraining the invariant masses of the three  $\gamma\gamma$ pairs and the $\pi^0\pi^0\pi^0$ system to the nominal $\pi^0$ and $\eta^\prime$ masses, respectively.
Events with $\chi^2_{\rm{8C}}<50$ are accepted for further analysis. If there is more than one combination in an event, the one with the smallest $\chi^2_{\rm{8C}}$ is retained.
To suppress possible background from the $J/\psi\to\gamma\eta\pi^0\pi^0$ decay, a seven-constraint (7C) kinematic fit is performed under the $\gamma\eta\pi^0\pi^0$ background hypothesis, and events with the probability of this 7C fit larger than that of the 8C signal hypothesis are discarded. 
In addition, events which have at least one $\gamma\gamma$ pair with invariant mass within the $\eta$ signal region, $|M(\gamma\gamma) - M_{\eta}| < 0.035$~ GeV$/c^2$, are rejected.
Possible background from the $J/\psi\to\omega\pi^0\pi^0$ decay is suppressed
by vetoing events with $|M(\gamma\pi^{0}) - M_{\omega}|<0.05$~ GeV$/c^{2}$,
where $M(\gamma\pi^{0})$ is the invariant mass of $\gamma\pi^{0}$ combination.

\begin{figure}[htbp]
  \centering
  \includegraphics[width=0.48\textwidth]{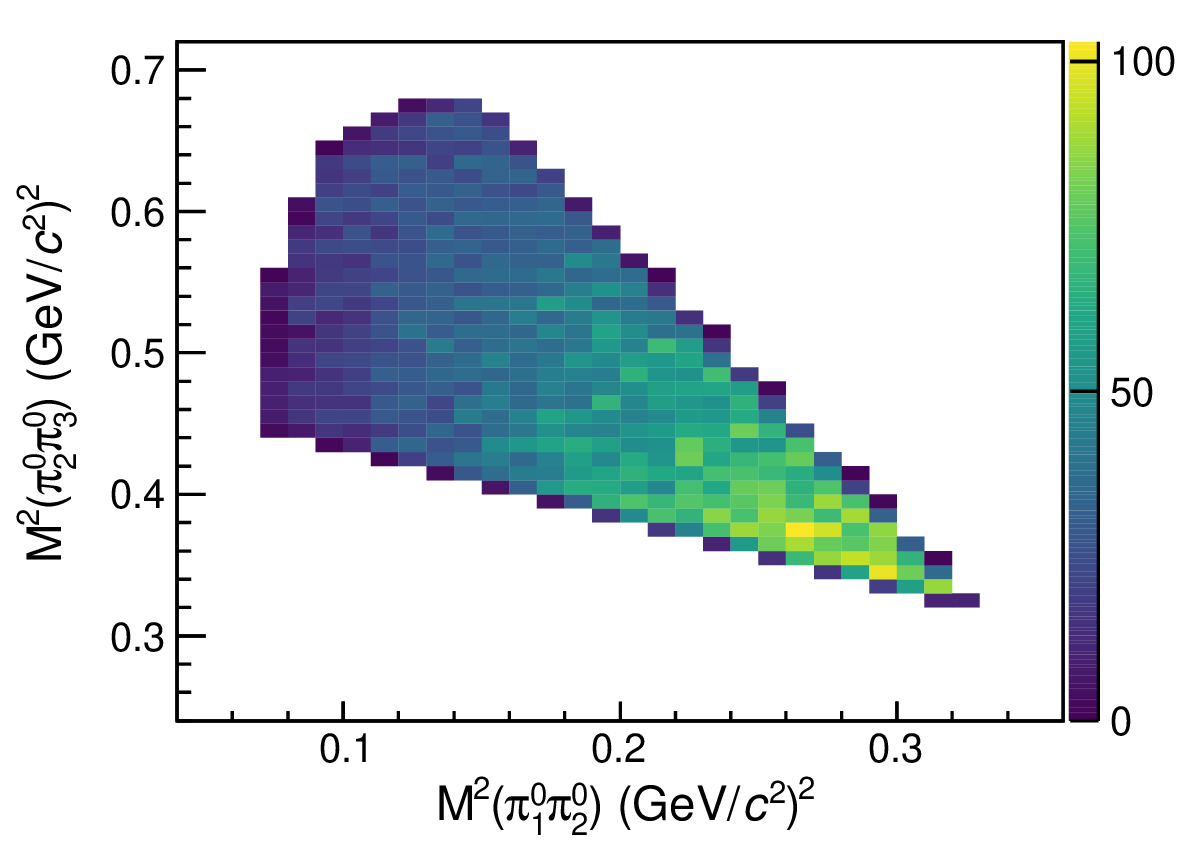}
 \vspace{-0.25cm}
  \caption{
     Dalitz plot for the $\eta^\prime \to \pi^0\pi^0\pi^0$ candidate events selected from data.  The $\pi^0_1$, $\pi^0_2$, and $\pi^0_3$ are ordered by descending energy in $\eta^\prime$ rest frame.
  }
  \label{fig:neu_mass}
\end{figure}

After applying the above selection criteria, the three $\pi^0$ candidates are ordered as $\pi^0_1$, $\pi^0_2$, and $\pi^0_3$, according to their descending energies in the $\eta^\prime$ rest frame. Figure~\ref{fig:neu_mass} presents the corresponding Dalitz plot for the selected data.  

\begin{figure}[htbp]
  \centering
  
  \includegraphics[width=0.48\textwidth]{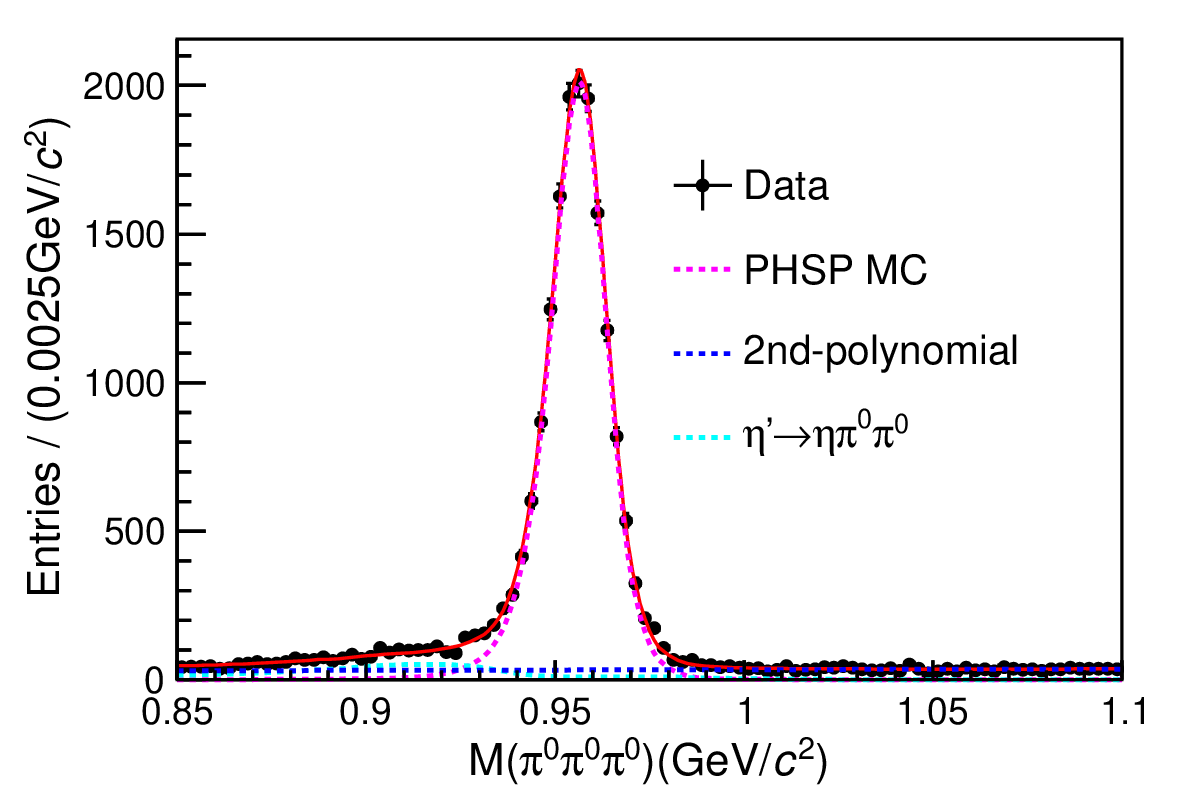}
  \caption{
  Invariant mass distribution of $\pi^0\pi^0\pi^0$ candidates without the $\eta^\prime$ mass constraint applied in the kinematic fit.
  }
  \label{fig:neu_fit_mass}
\end{figure}

The analysis of an MC sample of 10 billion $J/\psi$ inclusive events indicates that the background consists of a peaking component originating from the $J/\psi\to\gamma\eta^\prime$, with
$\eta^\prime\to\eta\pi^0\pi^0$, and a non-peaking background mainly coming from $J/\psi\to\gamma\pi^0\pi^0\pi^0$(the decay $J/\psi\to\pi^0\pi^0\pi^0\pi^0$ is forbidden). 
Using a dedicated MC sample with the decay amplitudes from Ref.~\cite{BESIII:2022tas}, the number of background events from the $\eta^\prime\to\eta\pi^0\pi^0$ is estimated to be $712\pm29$. While a non-peaking background will be described by a phase space distributed $J/\psi \to \gamma\pi^0\pi^0\pi^0$ MC simulation.
To estimate the intensity of non-peaking background, a 7C
kinematic fit without applying the constraint on the $\eta^\prime$ mass is performed. 
The obtained $M(\pi^0\pi^0\pi^0)$ distribution is shown in Fig.~\ref{fig:neu_fit_mass} and fitted with the signal shape convoluted with a Gaussian resolution function, simulated  $\eta^\prime\to\eta\pi^0\pi^0$ shape for peaking background and fixed to its expected intensity, and a non-peaking background described by a second-order polynomial function.  
The fit yields $15,368\pm129$ $\eta^\prime\to\pi^0\pi^0\pi^0$ signal events.
After taking into account the detection efficiencies with and without the $\eta^\prime$ mass constraint, the number of non-peaking background events in 8C sample is estimated to be $250\pm4$.
The comparison of the $\pi^0\pi^0$ mass spectra between the  $J/\psi\to\gamma\pi^0\pi^0\pi^0$ MC simulation and the events in the $\eta^\prime$ sideband regions, 
$(0.845,0.87)$ GeV/$c^2$ or $(1.02,1.045)$ GeV/$c^2$, 
indicates that the MC simulation accurately describes the non-peaking background.

\section{Amplitude analysis} 
\label{sec:amp}

An amplitude model based on the formalism of the isobar model~\cite{ref::isobar1,ref::isobar2} is established
to study the substructures of the $\eta^\prime\to \pi^{+}\pi^{-}\pi^{0}$ and $\eta^\prime\to \pi^{0}\pi^{0}\pi^{0}$ decays. 
In this model, 
the matrix element is parameterized by:
\begin{linenomath*}
\begin{equation}
\centering
\label{eq:matrix}
\mathcal{M} = \sum_{L=0}^{L_{\max}} Z_LF^LA_L,
\end{equation}
\end{linenomath*}
where $Z_L$ describes the angular distribution of the final state particles, $F^L$ is the barrier factor for the production of the partial wave, and the partial-wave amplitudes $A_L$ are $L$-dependent functions of the $\pi\pi$ squared invariant mass $s$. The sum is taken over the orbital angular momentum $L$ of the two-body partial waves.

As reported in the previous \text{BESIII} analysis~\cite{ref::kang}, contributions from $S$- and $P$-waves were observed in the $\eta^\prime\to\pi^+\pi^-\pi^0$ final state, which will be considered in this work and represented as:
\begin{linenomath*}
\begin{alignat}{2}
    A_0 &= c_{\rm{NR}}^C + c_{S}^C F_R^0 W_S(s), \notag \\
    A_1 &= c_{P^+} F_R^1 W_{P}(u)+ c_{P^-} F_R^1 W_{P}(t);
\end{alignat}
\end{linenomath*}
while in $\eta^\prime\to\pi^0\pi^0\pi^0$, due to $C$-parity limitation, only the $S$-wave ($L=0$) is considered:
\begin{linenomath*}
\begin{align}
    A_0(s,t,u) = c_{\rm{NR}}^N + c_{S}^N F_R^0 W_S.
\end{align}
\end{linenomath*}
The contribution of non-resonant (NR) decays is represented by $c_{\rm{NR}}=A_{\rm{NR}} e^{i\phi_{\rm{NR}}}$, which is fixed to unity in the fit.
The subscripts $R$ denote the intermediate resonances, while the superscripts $C$ and $N$ indicate the charged and neutral channels, respectively. The factor $c_{x}=A_{x} e^{i\phi_x}$ for a given intermediate state includes the amplitude $A_x$ and phase $\phi_x$, which are extracted via fitting to the data.  
The $\pi$-$\pi$ scattering amplitudes $W_L$ are described following the formalism from Ref.~\cite{PhysRevD.83.074004},
\begin{linenomath*}
\begin{equation}
    W_L(s) = \frac{1}{\cot \delta_{L}(s) - i},
\end{equation}
\end{linenomath*}
with
{\small
\begin{linenomath*}
\begin{align}
\cot \delta_{0}(s) &= \frac{\sqrt{s}}{2 p_{\pi}} \frac{M_{\pi}^{2}}{s - M_{\pi}^{2}/2}
\left( \frac{M_{\pi}}{\sqrt{s}} + B_0^S + B_1^S \omega_0(s) + B_2^S \omega_0^2(s)\right),  \notag \\
\cot \delta_{1}(s) &= \frac{\sqrt{s}}{2 p_{\pi}^{3}} (M_{\rho^0}^{2} - s)
\left( \frac{2 M_{\pi}^{3}}{M_{\rho^0}^{2}\sqrt{s}} + B_0^P + B_1^P \omega_1(s) \right).  \notag \\
\omega_L(s) &= \frac{\sqrt{s} - \sqrt{s_L - s}}{\sqrt{s} + \sqrt{s_L - s}} -1. \notag
\end{align}
\end{linenomath*}
} Here $s$ is the squared $\pi\pi$  invariant mass, $p_\pi=\sqrt{s/4-M_\pi^2}$, $\sqrt{s_0}=2M_K$ and $\sqrt{s_1}=1.05$~GeV. The masses $M_{\rho^0}$, $M_\pi$, and $M_K$ are fixed to the world average values taken from the PDG~\cite{pdg}, and $B_0^S$, $B_1^S$, $B_2^S$, $B_0^P$, and $B_1^P$ are free parameters extracted via fitting to the data.

Alternatively, the $\pi$-$\pi$ $P$-wave can be interpreted as the $\rho^0$ meson, which is generally described using the Gounaris-Sakurai Breit-Wigner (GSBW) parameterization, described as:
\begin{linenomath*}
\begin{equation}
W_{\rho^0}^{\rm{GS}}(s)=\frac{m_{\rho^0}^{2}\left(1+d \cdot \Gamma_{\rho^0} / m_{\rho^0}\right)}{m_{\rho^0}^{2}-s+f(s)-i m_{\rho^0} \Gamma_{\rho^0}(s)},
\end{equation}
\end{linenomath*}
with
\begin{linenomath*}
\begin{align}
d &= \frac{3}{\pi} \cdot \frac{m_{\pi}^{2}}{q_{\pi}^{2}(m_{\rho^0})}
    \ln \left( \frac{m_{\rho^0}+2 q_{\pi}(m_{\rho^0})}{2 m_{\pi}} \right) \notag\\
  &\quad + \frac{m_{\rho^0}}{2 \pi q_{\pi}(m_{\rho^0})}
    - \frac{m_{\pi}^{2} m_{\rho^0}}{\pi q_{\pi}^{3}(m_{\rho^0})}, \notag\\
f(s) &= \Gamma_{\rho^0} \cdot \frac{m_{\rho^0}^{2}}{q_{\pi}^{3}(m_{\rho^0})}
    \bigg[
    q_{\pi}^{2}(s) \cdot (h(s) - h(m_{\rho^0}^{2})) \notag\\
    &\quad + (m_{\rho^0}^{2} - s) \cdot q_{\pi}^{2}(m_{\rho^0})
    \cdot \left. \frac{\mathrm{d}  h}{\mathrm{d} s} \right|_{s = m_{\rho^0}^{2}}
    \bigg], \notag \\
\Gamma_{\rho^0}(s) &= \Gamma_{\rho^0} \left( \frac{q_{\pi}(s)}{q_{\pi}(m_{\rho^0})} \right)^3 \cdot \frac{m_{\rho^0}}{\sqrt{s}}, \notag \\
h(s) &= \frac{2}{\pi} \cdot \frac{q_{\pi}(s)}{\sqrt{s}} \ln \left( \frac{\sqrt{s} + 2 q_{\pi}(s)}{2 m_{\pi}} \right). \notag
\end{align}
\end{linenomath*}
Here $q_{\pi}(m_{\rho^0})$ is the momentum of $\pi$ in the rest frame with $M_{\pi\pi}=m_{\rho^0}$. The mass $m_\rho^0$ and width $\Gamma_\rho^0$ are free parameters.

To describe the event density distribution on the Dalitz plot, the probability density functions (PDF) $\mathcal{P}$ are described as:
\begin{linenomath*}
\begin{align}
\mathcal{P}(s, t) =\; &
f_{s} \frac{\left|\mathcal{M}\left(s, t\right)\right|^{2} \varepsilon\left(s, t\right)}{
\int_{\rm{DP}} \left|\mathcal{M}\left(s, t\right)\right|^{2} \varepsilon\left(s, t\right) \mathrm{d}s \mathrm{d}t} 
\nonumber \\
+ &
\sum_{k} f_{b_{k}} \frac{B_{k}\left(s, t\right)}{\int_{\rm{DP}} B_{k}\left(s, t \right) \mathrm{d}s \mathrm{d}t }.
\end{align}
\end{linenomath*}
Here $f_s$ and $f_{b_k}$ are the signal and background fractions in the event sample, under the constraint that $f_s + \sum_{k}f_{b_k} = 1$. $\mathcal{M}(s, t)$ is the decay amplitudes described in Eq.~\ref{eq:matrix}.
$\varepsilon(s, t)$ is the detection efficiency, estimated as the average efficiency within a given bin of $s$ and $t$ and derived from the dedicated MC simulation based on the previous \text{BESIII} measurement~\cite{ref::kang}.
$B_k(s,t)$ represents the shape of different backgrounds, described in detail above and estimated according to the MC simulation.
The integral over the full Dalitz plot range gives the normalization factor in the denominator.

The free parameters in the PDF are optimized with an unbinned maximum-likelihood fit by combining $\eta^\prime\to\pi^+\pi^-\pi^0$ and $\eta^\prime\to\pi^0\pi^0\pi^0$ events.
The fit minimizes the total negative log-likelihood (NLL) value:
\begin{linenomath*}
\begin{equation}
-\ln \mathcal{L}=-\sum_{i=1}^{N_{1}} \ln\mathcal{P}_{i}-\sum_{j=1}^{N_{2}} \ln \mathcal{P}_{j}^{\prime},
\end{equation}
\end{linenomath*}
where $\mathcal{P}_i$, $\mathcal{P}'_j$ are the PDFs for the $\eta^\prime\to\pi^+\pi^-\pi^0$ event $i$-th and the $\eta^\prime\to\pi^0\pi^0\pi^0$ event $j$-th, respectively. The sum runs over all accepted events.

\begin{table}[htbp]
\centering
\caption{Fitted parameters under the $\pi$-$\pi$ scattering and GSBW scheme, where the errors are statistical only.}
\label{tab:par}
\resizebox{0.45\textwidth}{!}{
\small
\begin{tabular}{l|cc|cc}\hline\hline
\multirow{2}{*}{Par} & \multicolumn{2}{c|}{$\pi$-$\pi$ Scatt.} & \multicolumn{2}{c}{GSBW} \\
& $\pi^+\pi^-\pi^0$ & $\pi^0\pi^0\pi^0$
& $\pi^+\pi^-\pi^0$ & $\pi^0\pi^0\pi^0$ \\\hline
$A_{\mathrm{NR}}$       & 1.0 & 1.0 & 1.0 & 1.0 \\
$\Phi_{\mathrm{NR}}$    & 0 & 0 & 0 & 0 \\
$A_{S}$                 & $1.4 \pm 0.1$ & $-0.6 \pm 0.1$ & $1.6 \pm 0.1$ & $0.6 \pm 0.1$ \\
$\Phi_{S}$              & $0.9 \pm 0.1$ & $4.0 \pm 0.1$  & $1.1 \pm 0.1$ & $1.2 \pm 0.1$ \\
$A_{P/\rho^\pm}$   & $7.5 \pm 0.5$ & -- & $-0.4 \pm 0.03$ & -- \\
$\Phi_{P/\rho^\pm}$ & $0.7 \pm 0.1$ & -- & $0.4 \pm 0.1$ & -- \\
$B_{0}^{p}$             & $0.3 \pm 0.001$ & -- & -- & -- \\
$B_{1}^{p}$             & $2.0 \pm 0.001$ & -- & -- & -- \\
$M_{\rho^0}$       & \multicolumn{2}{c|}{775.5(fixed)~MeV }& \multicolumn{2}{c}{$(777.4 \pm 2.9)$~MeV} \\
$\Gamma_{\rho^0}$  & \multicolumn{2}{c|}{--} & \multicolumn{2}{c}{$(139.9 \pm 4.7)$~ MeV} \\\hline
\multicolumn{1}{l|}{$B_{0}^{s}$} & \multicolumn{2}{c|}{$-9.0 \pm 1.6$} & \multicolumn{2}{c}{$-9.8 \pm 1.5$} \\
\multicolumn{1}{l|}{$B_{1}^{s}$} & \multicolumn{2}{c|}{$-87.5 \pm 5.7$} & \multicolumn{2}{c}{$-79.4 \pm 5.3$} \\
\multicolumn{1}{l|}{$B_{2}^{s}$} & \multicolumn{2}{c|}{$-118.8 \pm 9.7$} & \multicolumn{2}{c}{$-110.3 \pm 5.1$} \\\hline\hline
\end{tabular}
}
\end{table}

\section{Fit Results}

Compared with the previous \text{BESIII} analysis, no additional significant resonance is seen. Therefore, the $P$-wave, the resonant $S$-wave, and the phase space $S$-wave are taken as the baseline solution for the amplitude analysis. The statistical significance for each component is confirmed to be larger than $5\sigma$.

\begin{figure*}[htbp]   
  \includegraphics[width=0.25\textwidth]{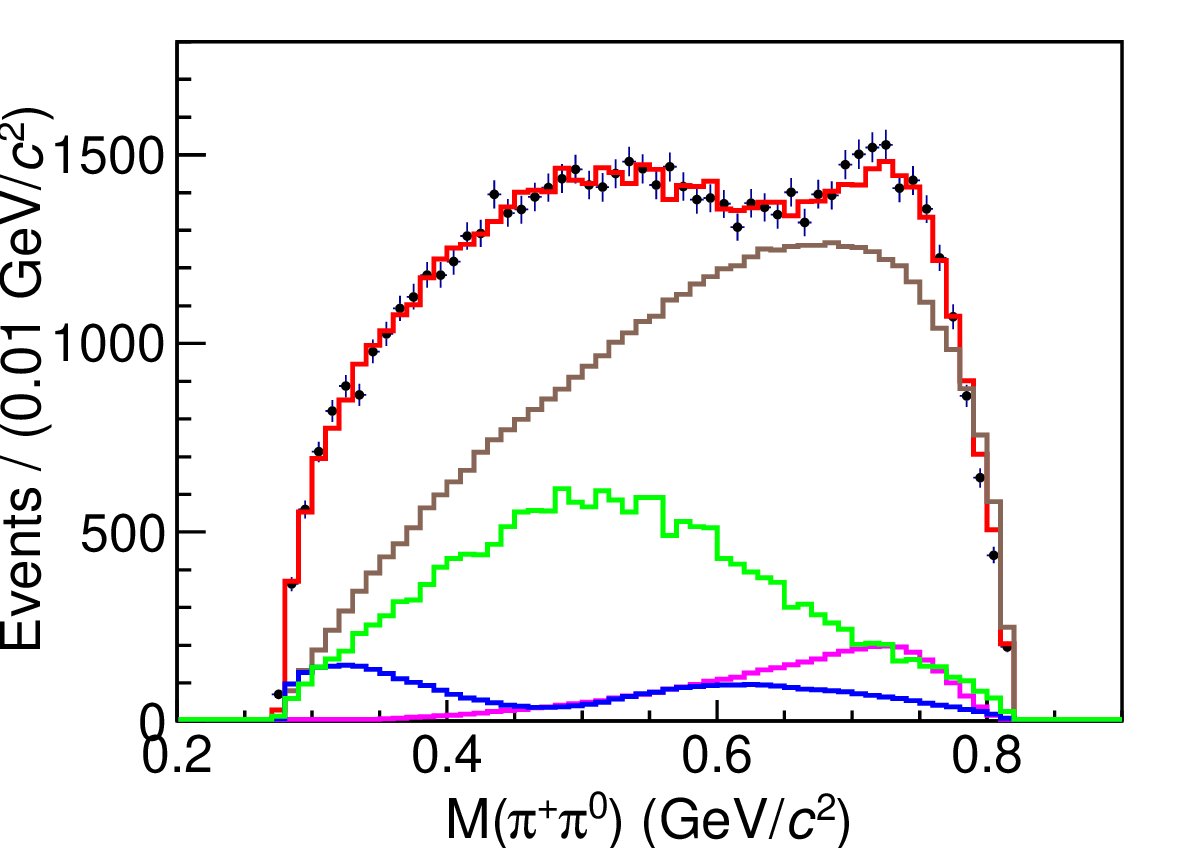} \put(-25,70){}
  \includegraphics[width=0.25\textwidth]{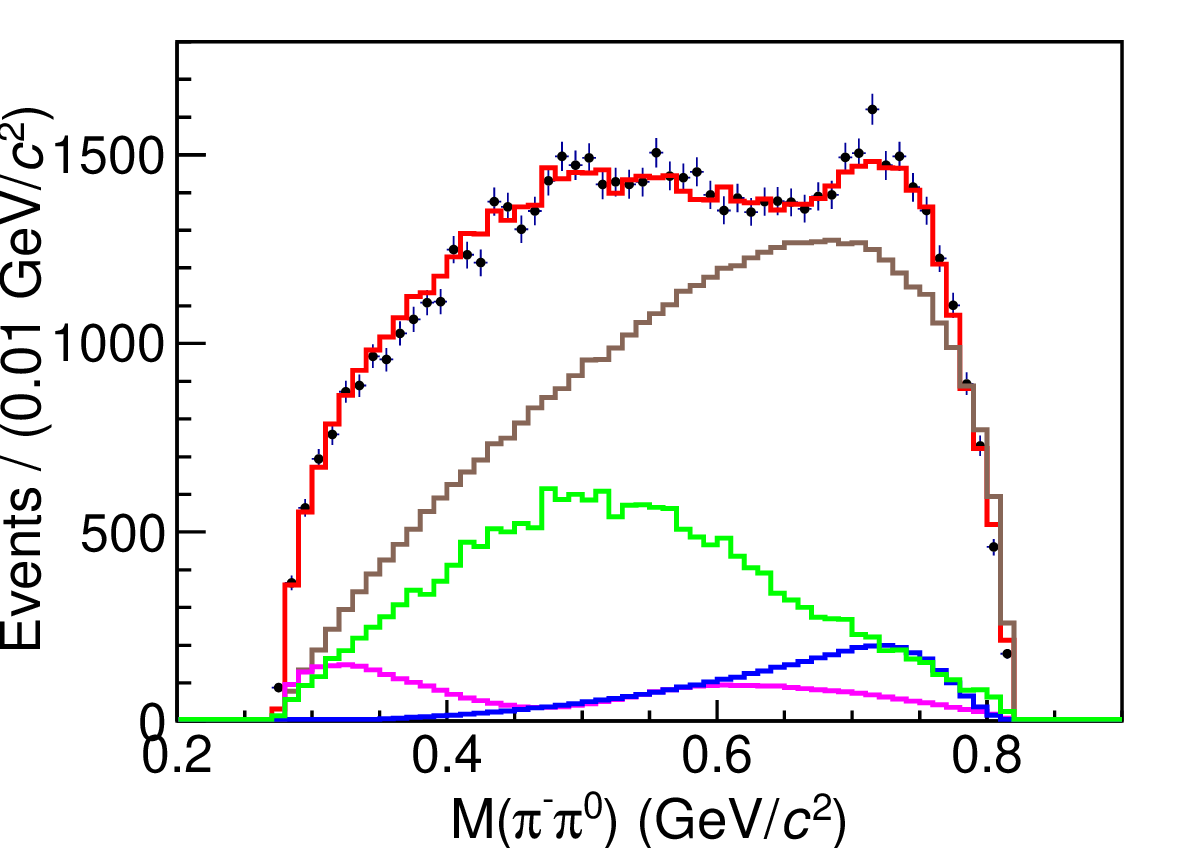} \put(-25,70){}
  \includegraphics[width=0.25\textwidth]{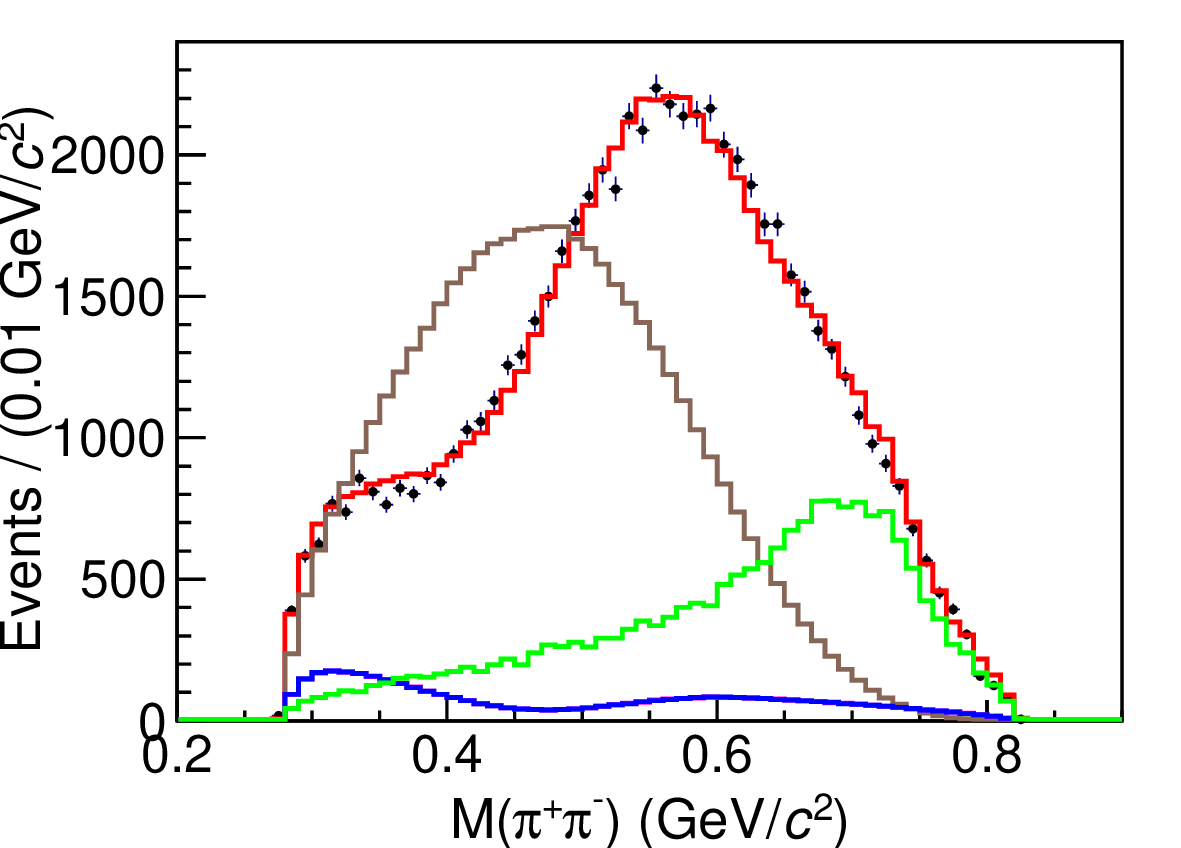} \put(-25,70){}
  \includegraphics[width=0.25\textwidth]{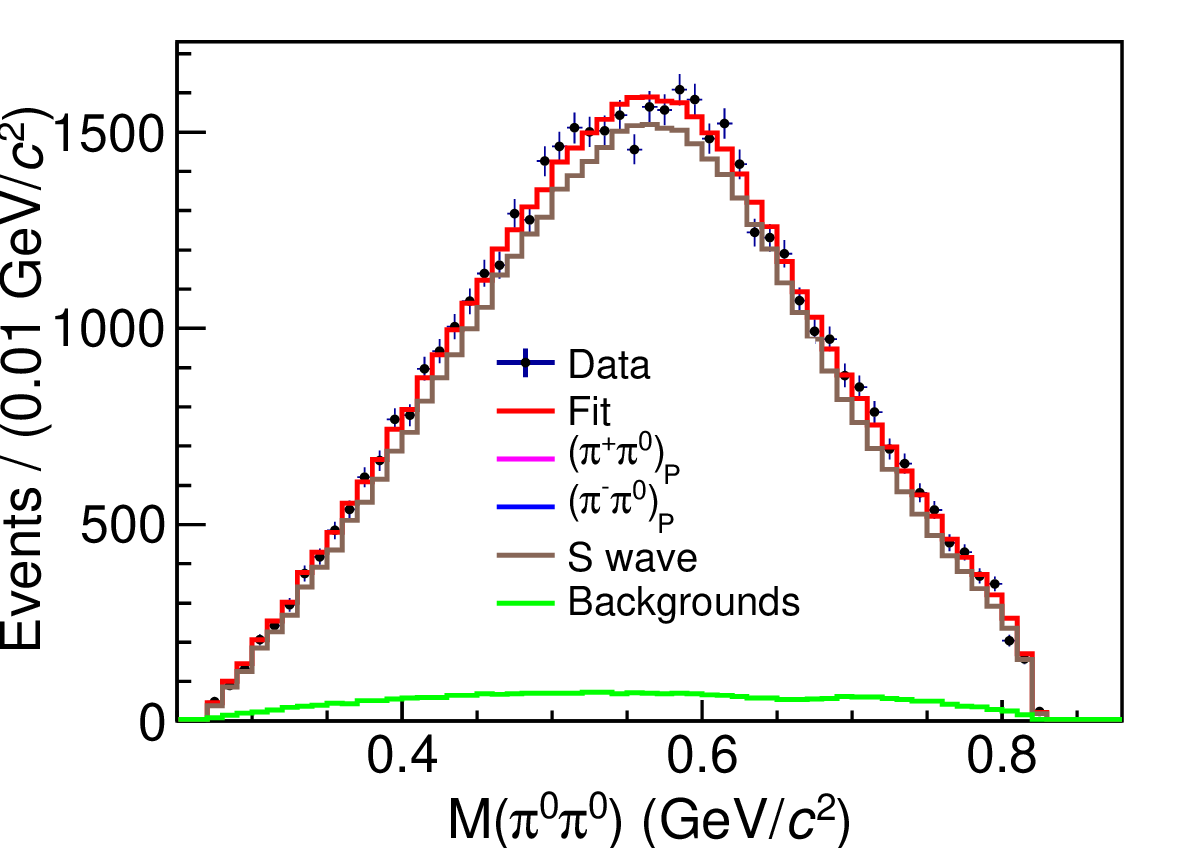} \put(-25,70){}\\
 
     \includegraphics[width=0.25\textwidth]{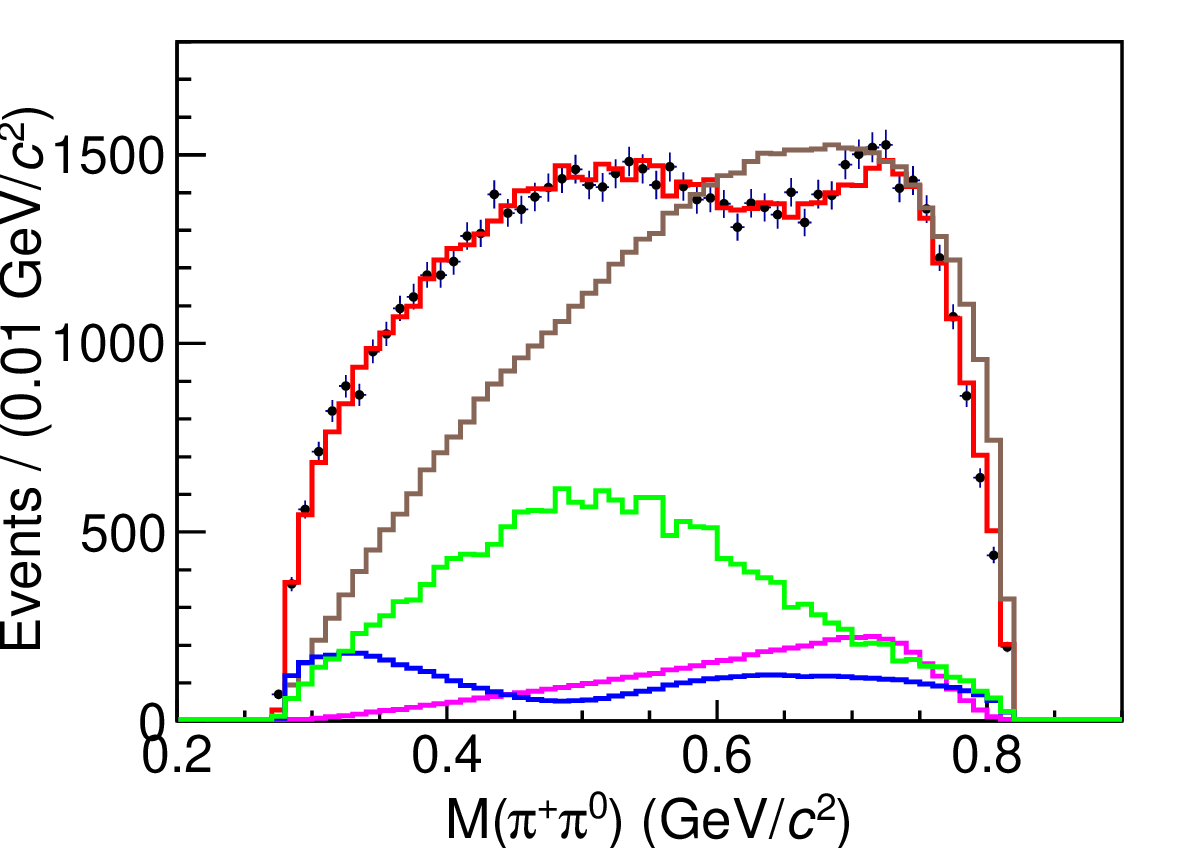} \put(-25,70){}
    \includegraphics[width=0.25\textwidth]{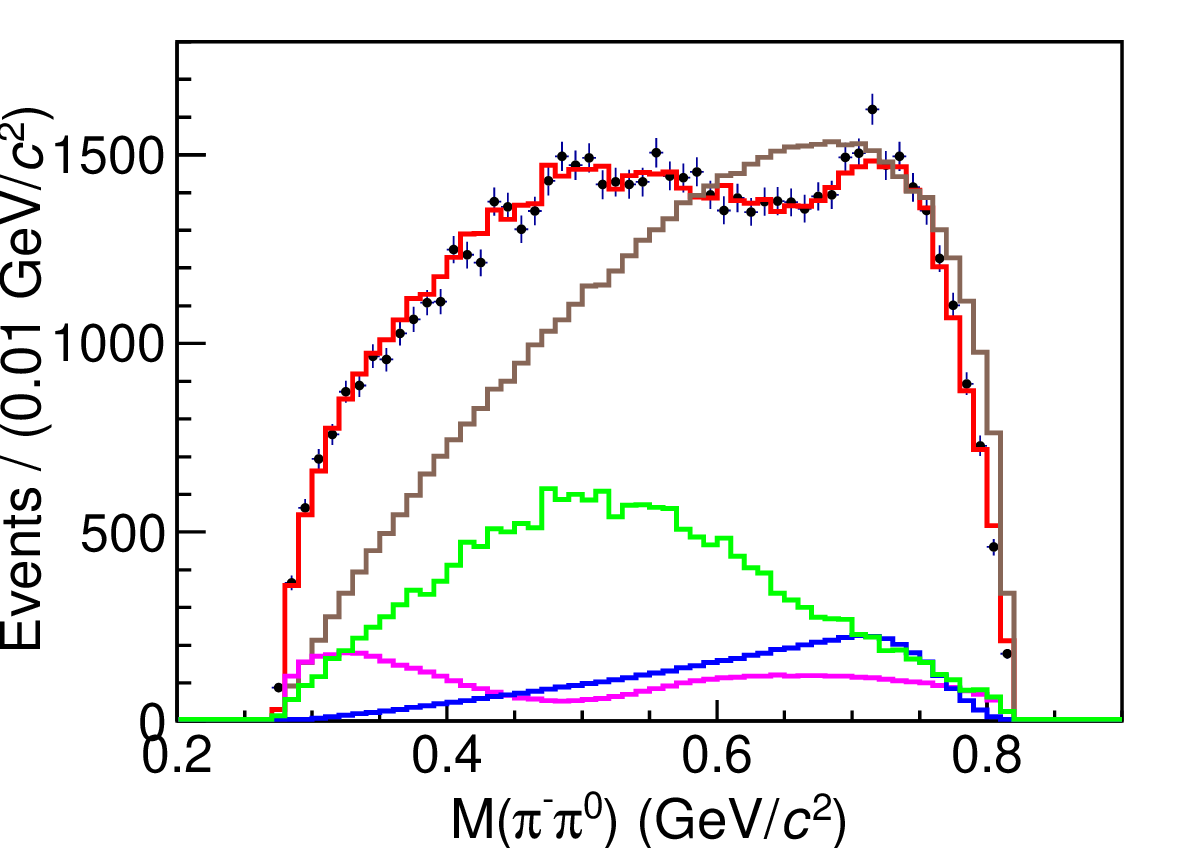} \put(-25,70){}
    \includegraphics[width=0.25\textwidth]{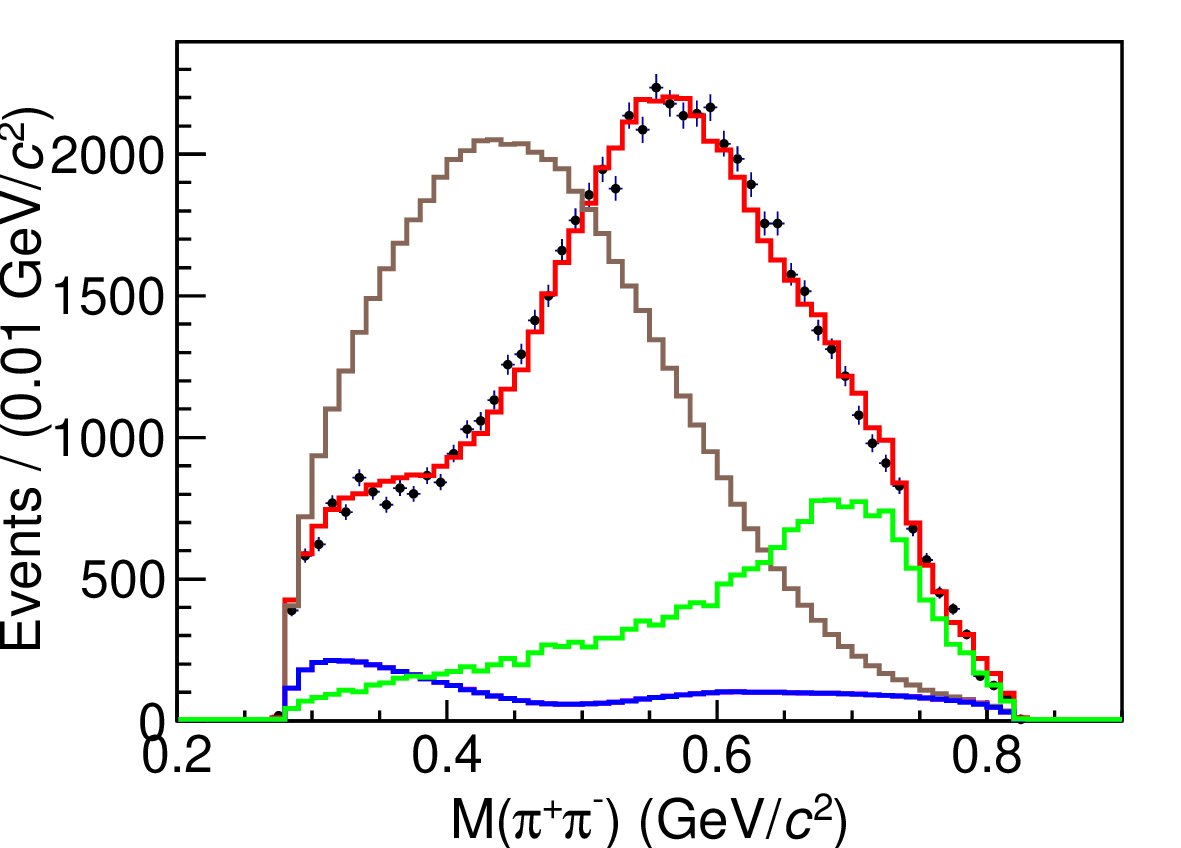} \put(-25,70){}
  \includegraphics[width=0.25\textwidth]{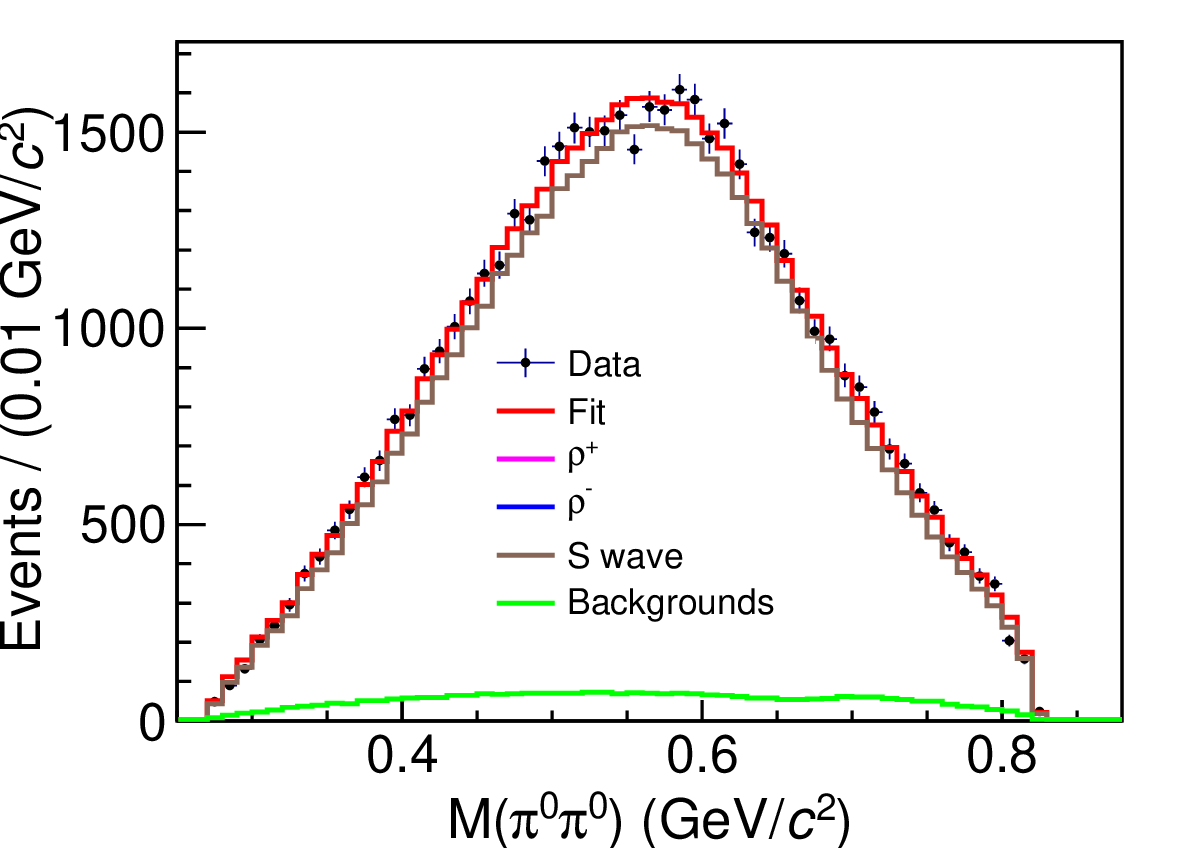} \put(-25,70){}

\caption{Projections of the invariant mass spectra for the $\pi^+\pi^0$, $\pi^+\pi^-$,  $\pi^-\pi^0$, and $\pi^0\pi^0$ combinations are shown. For the $\pi^0\pi^0$, 
all three possible $\pi^0\pi^0$ combinations are included. The distributions in the upper row are described with the $\pi$-$\pi$ scattering scheme, while those in the lower row are described with the GSBW scheme. Points with error bars represent data, and the solid red curves show the fit results for $\eta^\prime \to \pi^{+(0)}\pi^{-(0)}\pi^0$ decays. The pink, blue, brown, and green curves denote the $(\pi^+\pi^0)_P/\rho^{+}$, $(\pi^-\pi^0)_P/\rho^{-}$, $S$-wave, and background contributions, respectively.}
 
\label{fig:Pwave}
\end{figure*}

The best fit with $\pi$-$\pi$ scattering amplitudes yields an NLL value of $-5921.61$ with the free parameters listed in Table~\ref{tab:par}. The corresponding poles of the $\pi$-$\pi$ $S$- and $P$-waves are determined to be $(553.2\pm8.7)-i(220.0 \pm 8.0)$~MeV and $775.49 \mathrm{(fixed)}-i(60.7 \pm 0.2)$~MeV, respectively. 
The obtained masses and widths of the $\sigma$ and $\rho^0$ resonances are in reasonable agreement with the PDG values, even though the width of $\rho^0$ resonance is slightly lower.
A comparison between the data and fit projections on the $\pi\pi$ invariant mass is displayed in Fig.~\ref{fig:Pwave}.

The branching fraction for each component is determined by:
\begin{linenomath*}
\begin{equation}
\mathcal{B} = \frac{N_L}{N_{J/\psi} \cdot \mathcal{B}(J/\psi \to \gamma \eta^\prime) \cdot \mathcal{B}(\pi^0 \to \gamma \gamma)^n \cdot \varepsilon}.
\end{equation}
\end{linenomath*}
where $n=1$ for the charged mode and $n=3$ for the neutral mode. 
The signal yields $N_L$ defined as the integrals over the Dalitz plot of a single decay
amplitude squared, the detection efficiencies $\varepsilon$ obtained from the MC sample weighted with each amplitude, and the branching fractions for each component are summarized in Table~\ref{tab::branch}. In the calculation, the total number of $J/\psi$ events is taken from
Ref.~\cite{ref::Jpsi}, and the branching fractions for the  $J/\psi\to\gamma\eta^\prime$ and $\pi^0 \to \gamma \gamma$ decays are taken from the PDG~\cite{pdg}.
Because of the large interference between nonresonant and resonant $S$-waves, only the sum is used to describe the $S$-wave contribution.

An alternative fit with the GSBW model yields a slightly worse NLL value, $-5913.51$, with the same number of degrees of freedom. However, this fit also provides a good description of the data.
The obtained free parameters are also summarized in Table~\ref{tab:par}. 
Under the GSBW parameterization, the yield of $P$-wave is $30\%$ higher than that of the $\pi$-$\pi$ scattering scheme, and the yield of $S$-wave is $10\%$ higher. The selection efficiencies, however, are at the same level.
Therefore, the branching fractions of the $P$- and $S$-waves are also higher than those of the $\pi$-$\pi$ scattering scheme. The yields, detection efficiency, and branching fractions are summarized in Table
~\ref{tab::branch}. The difference on the branching fractions for $\eta^\prime\to\pi\pi\pi$ decay with respect to the $\pi$-$\pi$ scattering scheme will be taken as systematic uncertainties, while the results for individual components are reported separately.

Additional contributions from $f_0(980)$ and $\rho(1450)^{\pm}$ are checked and the statistical significances are found to be 0.4$\sigma$ and 0.7$\sigma$, respectively. In the alternative fits,
$f_0(980)$ is described by the Flatte function with the parameters fixed using values from Ref.~\cite{ref::f0980} and $\rho(1450)^{\pm}$ is modeled using a mass-dependent BW function.

\begin{table}[htbp]
\centering 
\caption{Yields with statistical errors, detection efficiencies, and branching fractions for the studied $\eta^\prime$ decay
modes.}
\label{tab::branch}
\resizebox{0.48\textwidth}{!}{
\begin{tabular}{ccccc}
\hline\hline
Scheme & Mode & $N_{L}$ & $\varepsilon(\%)$ & $\mathcal{B}\left(\times 10^{-4}\right)$  \\
\hline
 \multirow{4}{*}{$\pi$-$\pi$ Scatt.}  & $(\pi^+\pi^0)_P$ & $4010 \pm 105$ & $25.11 \pm 0.68$ & $3.06 \pm 0.08$  \\ 
& $(\pi^-\pi^0)_P$ & $4016 \pm 105$ & $25.14 \pm 0.68$ & $3.07 \pm 0.08$  \\
& $(\pi^\pm\pi^0)_P$ & $11324 \pm 299$ & $25.46 \pm 0.41$ & $8.52 \pm 0.22$  \\  
 & $(\pi^+\pi^-\pi^0)_S$ & $46935 \pm 325$ & $26.38 \pm 0.20$ & $34.10 \pm 0.24$  \\
\hline

\multirow{4}{*}{GSBW}  & $\rho^+$ & $5610 \pm 192$ & $24.49 \pm 0.57$ & $4.31 \pm 0.19$ \\ 
 & $\rho^-$ & $5611 \pm 191$ & $24.49 \pm 0.57$ & $4.32 \pm 0.19$  \\
 & $\rho^{\pm}$ & $16039 \pm 574$ & $25.06 \pm 0.34$ & $12.27 \pm 0.44$  \\ 
 & $(\pi^+\pi^-\pi^0)_S$ & $56521 \pm 623$ & $26.30 \pm 0.19$ & $41.19 \pm 0.45$  \\
\hline
 & $\pi^+\pi^-\pi^0$ & $46531 \pm 258$ & $25.28 \pm 0.20$ & $35.28 \pm 0.20$  \\ 
 & $\pi^0\pi^0\pi^0$ & $15368 \pm 129$ & $8.53 \pm 0.23$ & $35.39 \pm 0.30$  \\
\hline\hline
\end{tabular}
}
\end{table}

\section{Systematic Uncertainties}

Various sources of systematic uncertainties in the measurement of the branching fractions for $\eta^\prime \to \pi^0\pi^0\pi^0$ and $\eta^\prime \to \pi^+\pi^-\pi^0$ decays are investigated and summarized in Table~\ref{tab:syserr}. The total systematic uncertainties are given by the quadratic sum of the individual uncertainties, assuming all the sources are independent.

Differences between data and MC samples for the tracking efficiency of charged pions, $\gamma_{R}$ and $\pi^0$ reconstruction are investigated using the $J/\psi\to\pi^+\pi^-\pi^0$ control sample. A transverse momentum and cos$\theta$ dependent corrections to the tracking efficiency is obtained by comparing the efficiency between data and MC simulation, where $\theta$ is the polar angle of the track. Similarly, a momentum-dependent correction for $\pi^0$ reconstruction is obtained. Alternative fits are performed by considering the efficiency corrections for charged pions or $\pi^0$, and the differences in the branching fraction are assigned as the systematic uncertainty. For the $\gamma_R$, we take the systematic uncertainty to be $1\%$~\cite{ref:daq}.

To evaluate the uncertainty from the kinematic fit, the track helix parameters and photon energy resolutions are corrected in the MC samples to reduce the difference between data and MC simulation, as detailed in Ref.~\cite{ref:kine}. Alternative fits are performed with the corrected MC samples, and the differences in the branching fraction are assigned as the systematic uncertainty associated with the kinematic fit.

To investigate the uncertainties of the background determination, alternative fits are performed for each background component individually. 
The peaking backgrounds $\eta^\prime\to\gamma\pi^+\pi^-$, $\eta^\prime\to\gamma\rho^0(\gamma\pi^+\pi^-)$, and $\eta^\prime\to\eta\pi^0\pi^0$ are varied according to the uncertainties of the branching fraction for $J/\psi\to\gamma\eta^\prime$ and the corresponding cascade decays in the PDG~\cite{pdg}.  
To estimate the systematic uncertainty from background determination, alternative fits are performed by varying each background component individually. 
For the peaking backgrounds, the contributions from $\eta^\prime\to\gamma\pi^+\pi^-$ and $\eta^\prime\to\gamma\rho^0(\to\gamma\pi^+\pi^-)$ are varied according to the uncertainties of the corresponding branching fractions reported by the PDG~\cite{pdg}. Different selection criteria for vetoing $\omega$ and $\eta$ mesons are applied by varying the corresponding mass windows by $\pm 1$ MeV/$c^2$. Alternative fits are performed, and the differences in the branching fractions are taken as the corresponding uncertainties.

In addition, the uncertainties from the total number of $J/\psi$ events~\cite{ref::Jpsi} and the branching fraction of $J/\psi\to\gamma\eta^\prime$ are also considered.

\setlength{\tabcolsep}{2pt}
\renewcommand{\arraystretch}{1.0}

\begin{table}[htbp]
\centering
\caption{Summary of systematic uncertainties (\%) for the branching fractions. 
The first four columns correspond to the $P$- and $S$-wave components in the $\pi$-$\pi$ scattering and GSBW schemes, 
while the last two columns correspond to the branching fractions for $\eta^\prime\to\pi^+\pi^-\pi^0$ and $\eta^\prime\to \pi^0\pi^0\pi^0$.}

\resizebox{\columnwidth}{!}{
\begin{tabular}{lcc|cc|cc}
\hline\hline
\multirow{2}{*}[-0.4ex]{Source}
& \multicolumn{2}{c|}{$\pi$-$\pi$  Scatt.}
& \multicolumn{2}{c|}{GSBW}
& \multicolumn{2}{c}{} \\
& $P$-wave & $S$-wave
& ~~$\rho^{\pm}$ & $S$-wave
& \raisebox{1.2ex}{$\pi^+\pi^-\pi^0$}
& \raisebox{1.2ex}{$\pi^0\pi^0\pi^0$} \\
\hline
MDC tracking          & 1.3 & 0.2 & 1.2 & 0.4 & 0.4 & -- \\
$\pi^{0}$ selection   & 0.2& 0.2& 0.4 & 0.1 & 0.1 & 0.8 \\
Radiative $\gamma_R$    & 1.0 & 1.0 & 1.0 & 1.0 & 1.0  & 1.0 \\
Kinematic fit         & 1.8 & 0.1 & 0.7 & 0.5 & 0.1 & 0.5 \\
Background            & 1.4 & 1.0 & 2.5 & 1.1 & 1.7  & 0.1 \\
Veto $\omega$ meson        & 0.2 & 0.3 & 0.4 & 0.2 & 0.02  & 0.3 \\
Veto $\eta$ meson          & 0.3 & 0.4 & 0.5 & 0.4 & 0.5  & 0.3 \\
Number of $J/\psi$    & 0.4 & 0.4 & 0.4 & 0.4 & 0.4  & 0.4 \\
$\mathcal{B}(J/\psi\to\gamma\eta^\prime)$
                      & 1.3 & 1.3 & 1.3 & 1.3 & 1.3  & 1.3 \\

\hline
Total                 & 3.1 & 2.0 & 3.4 & 2.2 & 2.5  & 2.0 \\
\hline\hline
\end{tabular}
}
\label{tab:syserr}
\end{table}

\section{Summary}

Based on a data sample of $(10087\pm44)\times 10^6$ $J/\psi$ events collected with the \text{BESIII} detector, a combined amplitude analysis of $\eta^\prime\to\pi^+\pi^-\pi^0$ and $\eta^\prime\to\pi^0\pi^0\pi^0$ decays is performed. In addition to the nonresonant $S$-wave, significant $P$- and $S$-waves are confirmed in the $\eta^\prime\to\pi\pi\pi$ decays. 
The overall branching fractions of $\eta^\prime\to\pi^+\pi^-\pi^0$ and $\eta^\prime\to\pi^0\pi^0\pi^0$ are determined to be $(35.28\pm0.20\pm0.88)\times10^{-4}$ and $(35.39\pm0.30\pm0.71)\times10^{-4}$, respectively, which are in good agreement with and supersede the previous \text{BESIII} measurements~\cite{ref::kang} based on an earlier subset of the total \text{BESIII} data. 
However, the value of $\mathcal{B}(\eta^\prime\to\pi^0\pi^0\pi^0)$ is two times of the result from GAMS $(16\pm3.2)\times10^{-4}$, which is included in the PDG fitted value.

The $\pi$-$\pi$ scattering $S$ and $P$ phase shifts are used to describe the data. The pole position of the $P$-wave, $775.49\mathrm{(fixed)}-i(60.7\pm0.2)$~MeV, is consistent with the experimental measurements, although the width is smaller than the PDG average value.
The resonant $\pi$-$\pi$ $S$-wave, interpreted as the broad $\sigma$ meson, yields a pole at $(553.2\pm8.7)-i(220.0\pm8.0)$~MeV. Due to the large interference between nonresonant and resonant $S$-waves, only their sum is used to describe the $S$-wave contribution. The branching fractions with  $\pi$-$\pi$ scattering are determined to be
\begin{linenomath*}
\begin{align} 
\mathcal{B}(\eta^\prime\to\pi^+\pi^-\pi^0)_P=(8.52 \pm 0.22 \pm 0.26)\times10^{-4},\notag\\
\mathcal{B}(\eta^\prime\to\pi^+\pi^-\pi^0)_S=(34.10 \pm 0.24 \pm 0.68)\times10^{-4},\notag
\end{align}
\end{linenomath*}
which are consistent with and supersede the previous \text{BESIII} measurements based on the same scheme.
\begin{linenomath*}
\begin{align}
\small
\mathcal{B}(\eta^\prime\to\pi^+\pi^-\pi^0)_P=(7.44 \pm 0.60 \pm 1.26 \pm 1.84)\times10^{-4},\notag\\
\mathcal{B}(\eta^\prime\to\pi^+\pi^-\pi^0)_S=(37.63 \pm 0.77 \pm 2.22 \pm 4.48)\times10^{-4}.\notag
\end{align}
\end{linenomath*}
The first uncertainties are statistical, and the second are systematic, while the third is due to different propagator models in the amplitude analysis.

In addition, GSBW, which is generally used to describe the $\rho^\pm$ meson in the isobar model, could also provide a good description of the data. However, there are $30\%$ and $10\%$ differences in the $P$- and $S$- wave yields, respectively, between the two different schemes. The branching fractions under the GSBW scheme are determined to be
\begin{linenomath*}
\begin{align}
\mathcal{B}(\eta^\prime\to\pi^+\pi^-\pi^0)_P=(12.27\pm 0.44\pm 0.25)\times10^{-4},\notag\\
\mathcal{B}(\eta^\prime\to\pi^+\pi^-\pi^0)_S=(41.19 \pm 0.45\pm 0.91)\times10^{-4}.\notag
\end{align}
\end{linenomath*}
The pole position of the GSBW $S$-wave is $(514.7\pm8.6)-i(252.9\pm4.7)$~MeV.
Different models may help to understand the discrepancies between different theoretical works on these hadronic decays.

\section{ACKNOWLEDGMENTS}


The BESIII Collaboration thanks the staff of BEPCII (https://cstr.cn/31109.02.BEPC) and the IHEP computing center for their strong support. This work is supported in part by National Key R\&D Program of China under Contracts Nos. 2025YFA1613900, 2023YFA1606000, 2023YFA1606704; National Natural Science Foundation of China (NSFC) under Contracts Nos. 11635010, 11935015, 11935016, 11935018, 12025502, 12035009, 12035013, 12061131003, 12192260, 12192261, 12192262, 12192263, 12192264, 12192265, 12221005, 12225509, 12235017, 12342502, 12361141819, 12535005; the Chinese Academy of Sciences (CAS) Large-Scale Scientific Facility Program; the Strategic Priority Research Program of Chinese Academy of Sciences under Contract No. XDA0480600; CAS under Contract No. YSBR-101; 100 Talents Program of CAS; The Institute of Nuclear and Particle Physics (INPAC) and Shanghai Key Laboratory for Particle Physics and Cosmology; Agencia Nacional de Investigaci\'on y Desarrollo de Chile (ANID), Chile under Contract No. ANID CCTVal CIA250027; Istituto Nazionale di Fisica Nucleare, Italy; Knut and Alice Wallenberg Foundation under Contracts Nos. 2021.0174, 2021.0299, 2023.0315; Ministry of Development of Turkey under Contract No. DPT2006K-120470; National Research Foundation of Korea under Contract No. RS-2026-25486791; National Science and Technology fund of Mongolia; Polish National Science Centre under Contract No. 2024/53/B/ST2/00975; STFC (United Kingdom); Swedish Research Council under Contract No. 2019.04595; U. S. Department of Energy under Contract No. DE-FG02-05ER41374; Guangdong Basic and
Applied Basic Research Foundation 2024A1515012416.



\bibliographystyle{apsrev4-2}

\bibliography{pipi3gam.bib}

@article{Volkov:2022nuf,
    author = "Volkov, M. K. and Pivovarov, A. A. and Nurlan, K.",
    title = "{The decays {\ensuremath{\eta}}'{\textrightarrow} {\ensuremath{\pi}}{\ensuremath{\rho}} and {\ensuremath{\rho}} {\textrightarrow} {\ensuremath{\pi}}{\ensuremath{\eta}} in the chiral NJL model}",
    primaryClass = "hep-ph",
    doi = "10.1142/S0217732322501796",
    journal = "Mod. Phys. Lett. A",
    volume = "37",
    number = "27",
    pages = "2250179",
    year = "2022"
}

@article{Kubis:2009sb,
    author = "Kubis, Bastian and Schneider, Sebastian P.",
    title = "{The Cusp effect in eta-prime ---{\ensuremath{>}} eta pi pi decays}",
    reportNumber = "HISKP-TH-09-13",
    doi = "10.1140/epjc/s10052-009-1054-7",
    journal = "Eur. Phys. J. C",
    volume = "62",
    pages = "511--523",
    year = "2009"
}

@article{BESIII:2022tas,
    author = "Ablikim, Medina and others",
    collaboration = "BESIII Collaboration",
    title = "{Evidence for the Cusp Effect in {\ensuremath{\eta}}' Decays into {\ensuremath{\eta}}{\ensuremath{\pi}}0{\ensuremath{\pi}}0}",
    doi = "10.1103/PhysRevLett.130.081901",
    journal = "Phys. Rev. Lett.",
    volume = "130",
    number = "8",
    pages = "081901",
    year = "2023"
}

@article{Gross:1979ur,
    author = "Gross, David J. and Treiman, S. B. and Wilczek, Frank",
    title = "{Light Quark Masses and Isospin Violation}",
    reportNumber = "Print-79-0123 (PRINCETON)",
    doi = "10.1103/PhysRevD.19.2188",
    journal = "Phys. Rev. D",
    volume = "19",
    pages = "2188",
    year = "1979"
}

@article{ref::CLEO,
  author         = "Naik and others",
  collaboration  = "CLEO Collaboration",
  journal        = "Phys. Rev. Lett.",
  volume         = "102",
  pages          = "061801",
  year           = "2009",
  doi            = "10.1103/PhysRevLett.102.061801"
}

@article{ref::BESetap,
  author         = "M.~Ablikim and others",
  collaboration  = "BESIII Collaboration",
  title          = "{First observation of $\eta(1405) \to f_0(980)\pi^0$}",
  journal        = "Phys. Rev. Lett.",
  volume         = "108",
  pages          = "182001",
  year           = "2012",
  doi            = "10.1103/PhysRevLett.108.182001"
}

@article{ref::kang,
  author         = "M.~Ablikim and others",
  collaboration  = "BESIII Collaboration",
  journal        = "Phys. Rev. Lett.",
  volume         = "118",
  pages          = "012001",
  year           = "2017",
  doi            = "10.1103/PhysRevLett.118.012001"
}

@article{Ablikim:2009aa,
  author         = "M.~Ablikim and others",
  collaboration  = "BESIII Collaboration",
  journal        = "Nucl. Instrum. Meth. A",
  volume         = "614",
  pages          = "345--399",
  year           = "2010",
  doi            = "10.1016/j.nima.2009.12.050"
}

@inproceedings{Yu:IPAC2016-TUYA01,
  author       = "C.~H.~Yu and others",
  title        = "{BEPCII Performance and Beam Dynamics Studies on Luminosity}",
  booktitle    = "Proceedings of IPAC2016",
  year         = "2016",
  address      = "Busan, Korea",
  doi          = "10.18429/JACoW-IPAC2016-TUYA01"
}

@article{Ablikim:2019hff,
  author        = "M.~Ablikim and others",
  collaboration = "BESIII Collaboration",
  journal       = "Chin. Phys. C",
  volume        = "44",
  pages         = "040001",
  year          = "2020",
  doi           = "10.1088/1674-1137/44/4/040001"
}

@article{EcmsMea,
  author       = {J.~Lu and Y.~Xiao and X.~Ji},
  title        = "{Online monitoring of the center-of-mass energy from real data at BESIII}",
  journal      = {Radiat. Detect. Technol. Methods},
  volume       = {4},
  pages        = {337--344},
  year         = {2020},
  doi          = {10.1007/s41605-020-00188-8}
}

@article{EventFilter,
  author       = {J.~W.~Zhang and L.~H.~Wu and S.~S.~Sun and others},
  title        = "{Suppression of top-up injection backgrounds with offline event filter in the BESIII experiment}",
  journal      = {Radiat. Detect. Technol. Methods},
  volume       = {6},
  pages        = {289--293},
  year         = {2022},
  doi          = {10.1007/s41605-022-00331-7}
}

@article{Li:2017etof,
 author       = {X.~Li and others},
  title        = "{Study of MRPC technology for BESIII endcap-TOF upgrade}",
  journal      = {Radiat. Detect. Technol. Methods},
  volume       = {1},
  pages        = {13},
  year         = {2017},
  doi          = {10.1007/s41605-017-0014-2}
}

@article{Guo:2017etof,
  author       = {Y.~X.~Guo and others},
  title        = "{The study of time calibration for upgraded end-cap TOF of BESIII}",
  journal      = {Radiat. Detect. Technol. Methods},
  volume       = {1},
  pages        = {15},
  year         = {2017},
  doi          = {10.1007/s41605-017-0012-4}
}

@article{Cao:2020etof,
  author  = {P.~Cao and others},
  title   = {{Design and construction of the BESIII endcap TOF upgrade}},
  journal = {Nucl. Instrum. Meth. A},
  volume  = {953},
  pages   = {163053},
  year    = {2020},
  doi     = {10.1016/j.nima.2019.163053}
}

@article{geant4,
author       = {S.~Agostinelli and others},
  collaboration= {GEANT4 Collaboration},
  title        = "{GEANT4: A simulation toolkit}",
  journal      = "Nucl. Instrum. Meth. A",
  volume       = "506",
  pages        = "250--303",
  year         = "2003",
  doi          = "10.1016/S0168-9002(03)01368-8"
}

@article{kkmc1,
  author  = "S. Jadach and B.~F.~L. Ward and Z. Was",
  title   = "{Coherent exclusive exponentiation for precision Monte Carlo calculations}",
  journal = "Phys. Rev. D",
  volume  = "63",
  pages   = "113009",
  year    = "2001",
  doi     = "10.1103/PhysRevD.63.113009"
}

@article{kkmc2,
  author  = "S. Jadach and B.~F.~L. Ward and Z. Was",
  title   = "{The Precision Monte Carlo event generator KK for two fermion final states in $e^+e^-$ collisions}",
  journal = "Comput. Phys. Commun.",
  volume  = "130",
  pages   = "260--325",
  year    = "2000",
  doi     = "10.1016/S0010-4655(00)00048-5"
}

@article{evtgen1,
  author  = "D.~J.~Lange",
  title   = "{The EvtGen particle decay simulation package}",
  journal = "Nucl. Instrum. Meth. A",
  volume  = "462",
  pages   = "152--155",
  year    = "2001",
  doi     = "10.1016/S0168-9002(01)00089-4"
}

@article{evtgen2,
  author  = "R.~G.~Ping",
  title   = "{EvtGen: An Event generator for e+ e- collision experiments}",
  journal = "Chin. Phys. C",
  volume  = "32",
  pages   = "599",
  year    = "2008",
  doi     = "10.1088/1674-1137/32/8/001"
}

@article{pdg,
    author = "Navas, S. and others",
    collaboration = "Particle Data Group",
    title = "{Review of particle physics}",
    journal = "Phys. Rev. D",
    volume = "110",
    number = "3",
    pages = "030001",
    year = "2024",
    doi = "10.1103/PhysRevD.110.030001",
}

@article{lundcharm1,
  author  = "J.~C.~Chen and G.~S.~Huang and X.~R.~Qi and D.~H.~Zhang and Y.~S.~Zhu",
  title   = "{Event generator for J/psi and psi(2S) decays}",
  journal = "Phys. Rev. D",
  volume  = "62",
  pages   = "034003",
  year    = "2000",
  doi     = "10.1103/PhysRevD.62.034003"
}

@article{lundcharm2,
  author  = "R.~L.~Yang and R.~G.~Ping and H.~Chen",
  title   = "{Tuning and Validation of the Lundcharm Model with $J/\psi$ Decays}",
  journal = "Chin. Phys. Lett.",
  volume  = "31",
  pages   = "061301",
  year    = "2014",
  doi     = "10.1088/0256-307X/31/6/061301"
}

@article{photos2,
  author  = "E.~Barberio and B.~van Eijk and Z.~Was",
  title   = "{PHOTOS: A universal Monte Carlo for QED radiative corrections in decays}",
  journal = "Comput. Phys. Commun.",
  volume  = "66",
  pages   = "115--128",
  year    = "1991",
  doi     = "10.1016/0010-4655(91)90038-I"
}

@article{ref::rhogpp,
  author  = "G.~Toledo~S{\'a}nchez and J.~L.~Garcia-Luna and V.~Gonz{\'a}lez-Enciso",
  journal = "Phys. Rev. D",
  volume  = "76",
  pages   = "033001",
  year    = "2007",
  doi     = "10.1103/PhysRevD.76.033001"
}

@article{ref::f0980,
  author        = "M.~Ablikim and others",
  collaboration = "BESIII Collaboration",
  title         = "{Resonances in $J/\psi \to \phi \pi^+ \pi^-$ and $\phi K^+ K^-$}",
  journal       = "Phys. Lett. B",
  volume        = "607",
  pages         = "243--253",
  year          = "2005",
  doi           = "10.1016/j.physletb.2004.12.041"
}

@article{ref::chapeak,
     author        = "M.~Ablikim and others",
  collaboration = "BESIII Collaboration",
  title         = "{Precision Study of $\eta' \to \gamma \pi^+ \pi^-$ Decay Dynamics}",
  journal       = "Phys. Rev. Lett.",
  volume        = "120",
  pages         = "242003",
  year          = "2018",
  doi           = "10.1103/PhysRevLett.120.242003"
}

@article{ref::isobar1,
    author        = "R.~M.~Sternheimer and S.~J.~Lindenbaum",
  title         = "{Extension of the Isobaric Nucleon Model and Study of Production Processes Involving π Mesons}",
  journal       = "Phys. Rev.",
  volume        = "123",
  pages         = "333--376",
  year          = "1961",
  doi           = "10.1103/PhysRev.123.333"
}

@article{ref::isobar2,
    author        = "D.~Herndon and P.~S{\"o}ding and R.~J.~Cashmore",
  title         = "{Generalized Isobar Model Formalism}",
  journal       = "Phys. Rev. D",
  volume        = "11",
  pages         = "3165--3183",
  year          = "1975",
  doi           = "10.1103/PhysRevD.11.3165"
}

@article{ref::Jpsi,
  author        = "M.~Ablikim and others",
  collaboration = "BESIII Collaboration",
  title         = "{Number of $J/\psi$ events at BESIII}",
  journal       = "Chin. Phys. C",
  volume        = "46",
  pages         = "074001",
  year          = "2022",
  doi = "10.1088/1674-1137/ac5c2e"
}

@article{ref:kine,
    author        = "M.~Ablikim and others",
  collaboration = "BESIII Collaboration",
  title         = "{Observation of $\eta' \to \pi^+ \pi^- e^+ e^-$ and search for $\eta' \to \pi^+ \pi^- \mu^+ \mu^-$}",
  journal       = "Phys. Rev. D",
  volume        = "87",
  pages         = "012002",
  year          = "2013",
  doi           = "10.1103/PhysRevD.87.012002"
}

@article{PhysRevD.83.074004,
  title = {Pion-pion scattering amplitude. IV. Improved analysis with once subtracted Roy-like equations up to 1100 MeV},
  author = {Garc\'{\i}a-Mart\'{\i}n, R. and Kami\ifmmode \acute{n}\else \'{n}\fi{}ski, R. and Pel\'aez, J. R. and Ruiz de Elvira, J. and Yndur\'ain, F. J.},
  journal = {Phys. Rev. D},
  volume = {83},
  issue = {7},
  pages = {074004},
  numpages = {34},
  year = {2011},
  month = {Apr},
  publisher = {American Physical Society},
  doi = {10.1103/PhysRevD.83.074004},
  url = {https://link.aps.org/doi/10.1103/PhysRevD.83.074004}
}

@article{Borasoy:2005du,
    author = "Borasoy, B. and Nissler, R.",
    title = "{Hadronic eta and eta-prime decays}",
    doi = "10.1140/epja/i2005-10188-9",
    journal = "Eur. Phys. J. A",
    volume = "26",
    pages = "383--398",
    year = "2005"
}

@article{Borasoy:2006uv,
    author = "Borasoy, B. and Meissner, Ulf-G. and Nissler, R.",
    title = "{On the extraction of the quark mass ratio (m(d)- m(u)) / m(s) from Gamma(eta-prime ---{\ensuremath{>}} pi0 pi+ pi-) / Gamma(eta-prime ---{\ensuremath{>}} eta pi+ pi-)}",
    reportNumber = "HISKP-TH-06-24, FZJ-IKP-TH-2006-23",
    doi = "10.1016/j.physletb.2006.10.020",
    journal = "Phys. Lett. B",
    volume = "643",
    pages = "41--45",
    year = "2006"
}

@article{ref:daq,
  author  = {N. Berger and others},
  title   = {Trigger efficiencies at {BESIII}},
  journal = {Chin. Phys. C},
  volume  = {34},
  pages   = {1779--1784},
  year    = {2010},
  doi     = {10.1088/1674-1137/34/12/001},
}

@article{box1,
  author  = {Wess, J. and Zumino, B.},
  title   = {Consequences of anomalous Ward identities},
  journal = {Phys. Lett. B},
  volume  = {37},
  pages   = {95--97},
  year    = {1971},
  doi     = {10.1016/0370-2693(71)90582-X}
}

@article{box2,
  author  = {Witten, E.},
  title   = {Global Aspects of Current Algebra},
  journal = {Nucl. Phys. B},
  volume  = {223},
  pages   = {422--432},
  year    = {1983},
  doi     = {10.1016/0550-3213(83)90063-9}
}

\end{document}